\documentclass[a4paper,11pt]{article}
\pdfoutput=1 
\usepackage{jinstpub} 
\usepackage{placeins}

\title{\boldmath First Demonstration of the Use of LG-SiPMs for Optical Readout of a TPC}

\author[a, b]{A. Gola,}
\author[1, c]{K. Majumdar,\note{Corresponding author}}
\author[a, c]{G. Casse,}
\author[c]{K. Mavrokoridis,}
\author[a, b, d]{S. Merzi,}
\author[c]{and L. Parsons Franca}

\affiliation[a]{Fondazione Bruno Kessler (FBK), Via Sommarive, 18, 38123 Povo TN, IT}
\affiliation[b]{Trento Institute for Fundamental Physics and Applications (TIFPA), \\ Via Sommarive, 14, 38123 Povo TN, IT}
\affiliation[c]{University of Liverpool, Dept. of Physics, Oliver Lodge Bld, Oxford Street, Liverpool, L69 7ZE, UK}
\affiliation[d]{University of Trento, Dept. of Physics, Via Sommarive, 14, 38123 Povo TN, IT}

\emailAdd{majumdar@liverpool.ac.uk}

\abstract{\\This paper describes a new method for optical readout of Time Projection Chambers (TPCs), based on the Linearly Graded Silicon Photomultiplier (LG-SiPM). This is a single photon-sensitive detector with excellent timing and 2D position resolution developed at Fondazione Bruno Kessler, Trento (FBK). The LG-SiPM produces time-varying voltage signals that are used to reconstruct the 3D position and energy of ionisation tracks generated inside the TPC.

The TPC used in this work contained room-temperature CF$_4$ gas at a pressure of 100~mbar, with two THGEMs to produce secondary scintillation light. A collimated $^{241}$Am source (Q$_\alpha$ = 5.486~MeV) was used to produce the ionisation tracks. The successful reconstruction of these tracks is demonstrated, and the consistency of the methodology characterised through varying the geometry of the tracks within the TPC. Energy reconstruction and deposition studies are also described, demonstrating the feasibility of the LG-SiPM as a potential option for optical TPC readout.}


\arxivnumber{2009.05086} 

\begin{document}

\maketitle

\FloatBarrier
\section{Introduction}

One of the most successful and widely used types of particle detector over the past few decades has been the Time Projection Chamber (TPC). This type of detector can be operated in a number of different ways - for example, using a gaseous or liquid scintillator medium, or a combination of both. One of the most commonly used media in such detectors is liquid argon (LAr). There are a number of LAr-based experiments currently operating around the world - for example, the three detectors of the Short Baseline Neutrino Program~\cite{SBNP}: SBND (which has 112~tons of LAr as its active TPC volume), MicroBooNE (89~tons) and ICARUS-T600 (470~tons), as well as the single- (411~tons) and dual-phase (300~tons) ProtoDUNE experiments \cite{ProtoDUNE-SP, ProtoDUNE-DP}. TPCs will only continue to increase in size and sophistication in the near future, with the DUNE project proposing the use of four 17,000 ton LArTPCs \cite{DuneCDRVol1, DuneCDRVol2, DuneCDRVol3, DuneCDRVol4, DuneIDRVol1, DuneIDRVol2, DuneIDRVol3} and a 20,000 ton LArTPC being operated at the DarkSide-20k experiment \cite{Darkside-20k}.
\\
\\
In a typical TPC, ionisation of the active scintillator material occurs as a particle travels through, producing scintillation light and free electrons. A micropattern charge amplifier, such as a Thick Gas Electron Multiplier (THGEM), is typically used to multiply the electrons to increase the signal-to-noise ratio, and the charge signal is then usually collected on a segmented anode plane. The electron multiplication also acts a secondary source of scintillation light \cite{THGEM_Review}.

Due to the single-photon sensitivity of many modern optical imaging devices, observing the secondary scintillation light is possible even at very low particle energies and signal-to-noise ratios, and can therefore offer lower energy thresholds than are possible using charge readouts. Depending on the design and nature of the optical readout device(s), detector construction and operation may also be simplified considerably, leading to reductions in costs through the life of an experiment.

Secondary scintillation light has previously been directly imaged using a range of devices: Geiger-mode avalanche photodiodes \cite{BondarGAPD}, silicon photomultipliers \cite{THGEMReadout} and standard CCDs \cite{ccdargon}. More recently, the use of EMCCDs and Timepix3-based cameras have been successfully demonstrated by the ARIADNE Project at the University of Liverpool \cite{emccdargon, ARIADNE_TDR, TPX_CF4}. This R\&D program is specifically dedicated to developing innovative optical readout systems for use in current and future dual-phase LArTPCs. 
\\
\\
In this paper, we introduce a new photo-detector: the \textbf{Linearly Graded Silicon Photomultiplier (LG-SiPM)}, along with a LG-SiPM-based camera system that can be used for the optical readout of TPCs that employ electron multiplication devices. This is a relatively new type of SiPM that offers position sensitivity along both the $x$ and $y$ axes. It is also expected to provide excellent timing performance, and therefore high resolution along the $z$ axis.

\FloatBarrier
\section{Hardware Description}

\subsection{The Liverpool 40l TPC}
\label{subsec:tpc_intro}

\begin{figure}[ht]
\centering
\includegraphics[width=\textwidth]{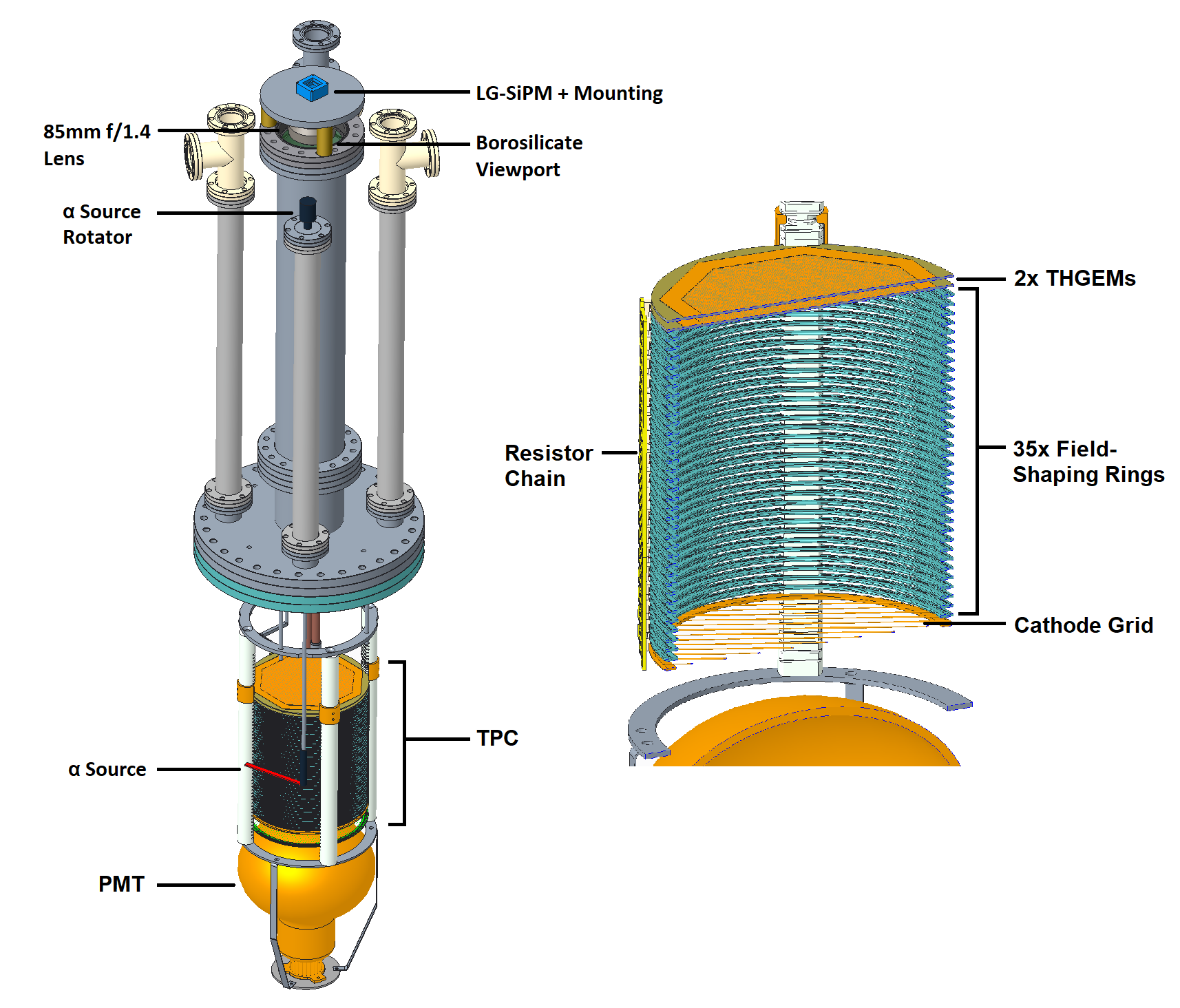}
\caption{\label{fig:40lDetector}(Left) A 3D model of the Liverpool 40l TPC, with key components labelled.  For clarity, the outer wall of the vacuum chamber has been omitted. (Right) A closeup view of the TPC, with key components labelled. The THGEMs are separated by 4~mm, as are the bottom THGEM and the top-most field-shaping ring.}
\end{figure}

The Liverpool 40l TPC was used for these studies, and is shown in Figure~\ref{fig:40lDetector}. The 19~cm tall and 17~cm diameter cylindrical field cage consists of 35 field-shaping rings, with a cathode grid located below them. The TPC volume was filled with pure CF$_4$ gas at a pressure of 100mbar, and the cathode and top-most field ring were biased to -3.0~kV and -1.1~kV respectively, giving an electric field of 100~V/cm within the TPC's drift volume. A single 8~inch diameter Hamamatsu R5912-20-MOD PMT is positioned below the cathode grid, and is used to detect the primary scintillation light for event timing and triggering purposes. It also acts as an additional detector for the secondary scintillation light produced in the THGEMs.

A collimated $^{241}$Am $\alpha$ source (Q$_\alpha$ = 5.486~MeV) is mounted on a rotating arm located 10~cm below the top-most field ring. A rotation angle of 0{\textdegree} corresponds to the arm pointing at a tangent to the field rings, and a maximum angle of 85{\textdegree} is possible. The source itself emits particles at an angle of 45{\textdegree} downwards.

Two identical THGEMs (Thick Gaseous Electron Multipliers) are located 4~mm above the top-most field ring, with a separation of 4~mm between them. Each THGEM consists of a 1~mm thick sheet of FR4 with copper layers on the top and bottom planes. Approximately 23,000 evenly spaced holes, each with a diameter of 500~$\mu$m and pitch of 800~$\mu$m, are drilled through all three layers, covering an area of 150~cm$^2$ per THGEM. Each hole has an additional 50~$\mu$m etched rim which serves to increase the THGEM's overall breakdown voltage. Each THGEM has an overall optical transparency of 35\%. The copper planes of each THGEM can be held at different voltages, as each of them is biased from its own independent power supply.  In these studies, both THGEMs were operated in the ``electron multiplication'' bias regime (described in more detail in Section~\ref{subsec:tpc_detection}): the bottom THGEM's bottom and top planes have been held at -0.85~kV and 0~V respectively, with the top THGEM at 0~V (bottom plane) and +0.85~kV (top plane). Holding both of the inner planes of the THGEMs at 0~V allows for larger absolute biases to be used on the outer planes without increasing the risk of discharges, leading to increased electron multiplication and secondary scintillation light production, and more stable operation.
\\
\\
The entire internal TPC assembly described above sits inside the 40l ultra-high-vacuum chamber, which is capped by a 33.6~cm diameter conflat (CF) flange.  Attached to and through this flange are the various electrical feedthroughs for the field cage and THGEM voltages, the fluid in/outlet for filling and evacuating the chamber, the rotation arm for the $\alpha$ source, as well as a 10~cm diameter pipe, at the top of which is a borosilicate viewport. The viewport is 1~m from the top THGEM plane.

Above this viewport is the hardware for optical readout of the TPC: a Canon 85~mm f/1.4 camera lens coupled to the LG-SiPM mounting using a standard C-mount threaded ring. The mounting itself has been designed at the University of Liverpool and 3D-printed from ABS+, allowing the LG-SiPM to be held securely and stably as close as possible to the lens.

\subsection{TPC Operation Principle}
\label{subsec:tpc_detection}

\begin{figure}[ht]
\centering
\includegraphics[trim={4.5cm 1.5cm 5.5cm 2cm}, clip, width=\textwidth]{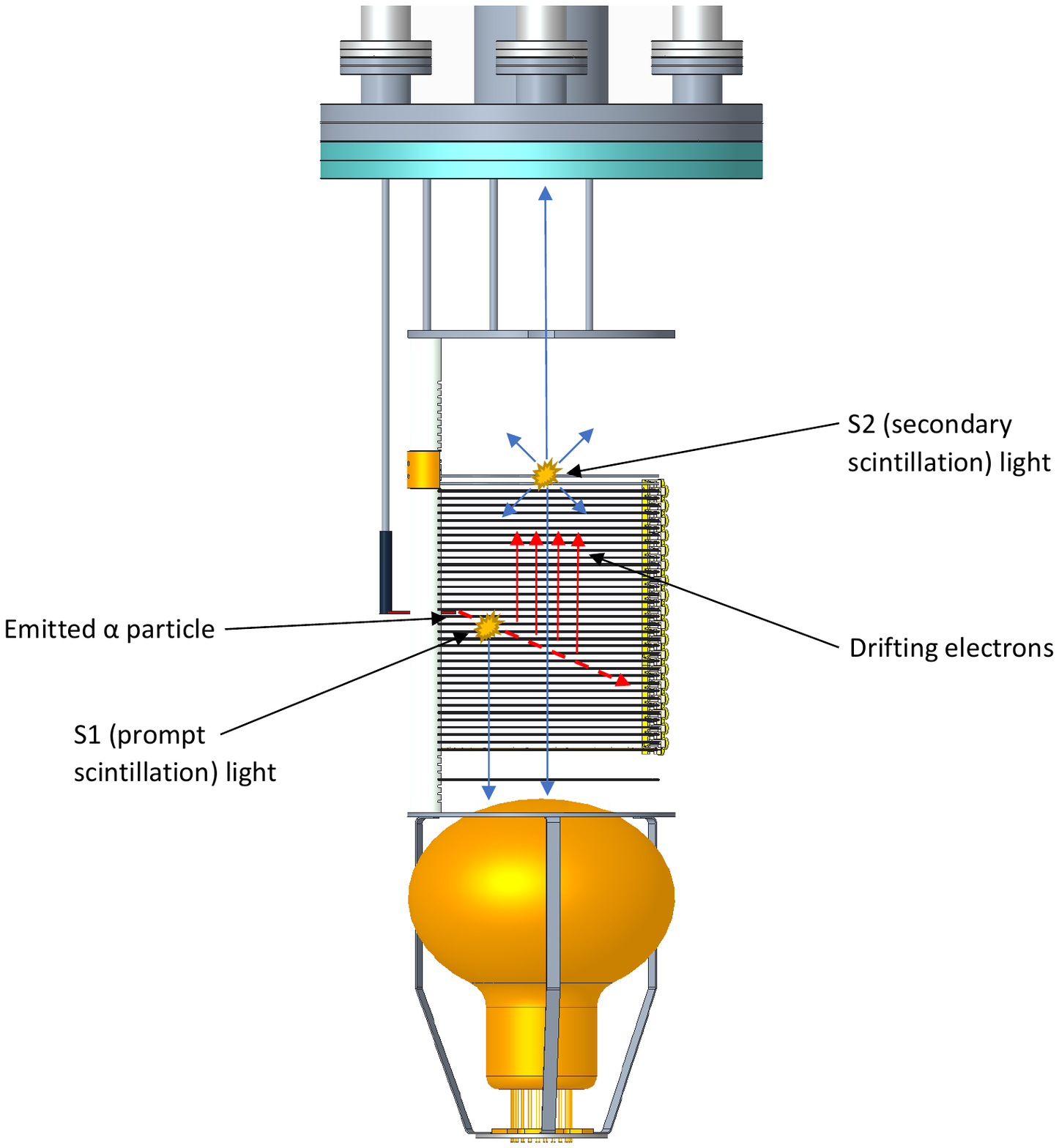}
\caption{\label{fig:DetectionPrinciple}Schematic view of the operation principle of the Liverpool 40l TPC. Details are given in the main text. Some detector components have been omitted from this model for clarity.}
\end{figure}

Figure~\ref{fig:DetectionPrinciple} shows the operating principle of the Liverpool 40l TPC. The key steps are discussed in the main text below.

\noindent When a particle enters the CF$_4$ volume, it immediately produces prompt scintillation light (S1), which is detected effectively instantaneously by the PMT. This is therefore the known start time of the event.  The particle also creates a track of ionised ions and free electrons as it moves through the volume. Due to the electric field within the TPC field cage, these electrons drift upwards towards the THGEMs.

The high potential difference across each THGEM causes the electrons to be accelerated through the THGEM holes. The exact bias across the THGEM determines the outcome of this process. At low biases, the electrons simply excite the gas molecules as they pass, and the subsequent de-excitations produce secondary (S2) light in an amount linearly proportional to the number of incident electrons. At higher biases, the electrons have enough energy to ionise the gas within the THGEM holes, causing avalanches (i.e. electron multiplication), and the amount of S2 light becomes exponentially proportional to the number of incident electrons. In a multi-THGEM setup such as that used in these studies, the total amount of S2 light produced is also related to the number of THGEMs (i.e. the number of multiplication stages) as well as the efficiency of transferring the electrons between the THGEMs. If the bias across a single THGEM is too high, there may be enough ionisation within a hole to form a conductive plasma, leading to a discharge between the planes.

The S2 light from each THGEM is emitted isotropically. Due to the previously noted optical transparency of the THGEMs, the PMT at the bottom of the detector will detect a large fraction of the downwards-emitted S2 light from the bottom THGEM, but only a small fraction of the light emitted in a downwards direction from the top THGEM. On the other hand, the majority of the S2 light detected by the LG-SiPM at the top of the viewport will have been emitted in an upwards direction from the top THGEM, with a minority contribution originating from the bottom THGEM. Due to the very small gap between the THGEMs, the separate S2 light pulses from each device cannot be individually resolved, but instead form a single pulse on both the PMT and LG-SiPM.

\subsection{The LG-SiPM}
\label{subsec:lgSiPM_intro}

The Linearly Graded Silicon Photomultiplier (LG-SiPM) has been designed as a new type of position-sensitive SiPM \cite{Gola2013}. When a photon hits the sensor's active area, the current generated by the SiPM microcells is split into four outputs, from which it is possible to calculate the photon's \textit{x} and \textit{y} coordinates, down to a theoretical spatial resolution equal to the size of the microcells - on the order of 30~$\mu$m.  The LG-SiPM has a fast time response - typically on the order of a few tens of ns, and more recent SiPM designs show the possibility of reducing this response time down to less than 5~ns \cite{Acerbi2018}.
\\
\\
The typical application of the LG-SiPM is for scintillation light readout in ultra-high-resolution Positron Emission Tomography (PET). Millimeter \cite{Berneking2018} and sub-millimeter \cite{Du2018} spatial resolutions have been recently demonstrated using $7.75 \times 7.75$~mm and $16 \times 16$~mm devices respectively. These studies also show the capability of the LG-SiPM in determining the precise position of a spatially localised, short duration (on the order of tens of ns) source of light.

The LG-SiPM's characteristics make it a promising photo-detector in a variety of other applications. The device is not a pixel detector, and so the signal produced by the four outputs are continuous in time. Therefore, if the active area is illuminated with a light source that changes position over time, the output signals of the LG-SiPM are expected to continuously change accordingly, and the equations of motion of the light source as it moves across the LG-SiPM's active area can be determined. This is of great interest in applications requiring light tracking with high sensitivity and high speed.

\begin{figure}[ht]
\centering
\vspace{3mm}
\includegraphics[width=0.75\textwidth]{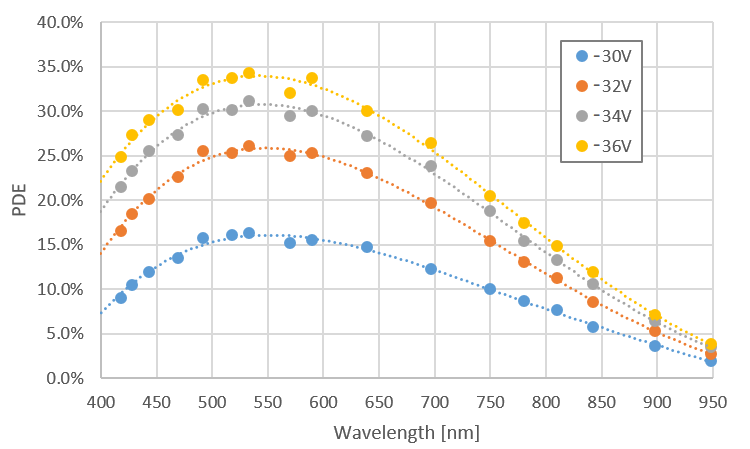}
\caption{\label{fig:LG-PDE}The photon detection efficiency (PDE) spectrum of the LG-SiPM as a function of the wavelength, and taken at different values of the supplied bias voltage.}
\end{figure}

At the same time, the Geiger mode operation of the LG-SiPM offers a high internal gain - on the order of a few million units of charge per primary photoelectron. This allows it to reliably detect single photons without the need for an additional multiplication stage.
\\
\\
The photon detection efficiency (PDE) spectrum of the LG-SiPM is shown in Figure~\ref{fig:LG-PDE}. Although there is a dependence of the absolute PDE on the supplied bias voltage (discussed in Section~\ref{subsec:lgSiPM_application}), it generally shows a peak at $\approx 550$~nm. In comparison, the secondary scintillation spectrum of CF$_4$ peaks at $\approx 625$~nm \cite{cf4Spectrum1, cf4Spectrum2}. At this wavelength, the LG-SiPM is seen to have a PDE of between 25 and 30\%.

\subsection{LG-SiPM Internal Structure}
\label{subsec:lgSiPM_structure}

\begin{figure}[ht]
\centering
\vspace{3mm}
\includegraphics[width=0.75\textwidth]{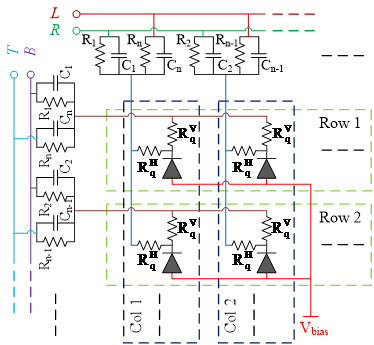}
\caption{\label{fig:SiPM-Circuit}Schematic of the microcell-level circuitry of the LG-SiPM. A full description is given in the main text.}
\end{figure}

A schematic of the LG-SiPM microcell circuitry is shown in Figure~\ref{fig:SiPM-Circuit}. Each microcell contains two quenching resistors, $R_q^H$ and $R_q^V$, connected to horizontal and vertical buses respectively (depicted as brown and light blue lines on the schematic). Each bus is terminated with a resistive current divider composed of two resistors, the values of which are dependent on the bus location and orientation. For example, the resistors $R_l$ and $R_{n-l-1}$ terminate the vertical bus at column $l$ (out of $n$ columns in total), and the horizontal bus at row $t$ out of $m$ total rows is terminated by resistors $R_t$ and $R_{m-t-1}$. The other ends of the resistors are connected to the two readout pads for each axis: denoted as $L$ and $R$ for the horizontal axis, and $T$ and $B$ for the vertical. Each microcell requires a bias voltage, $V_{bias}$, which is provided to the entire structure from an external source.

In the case of the vertical buses, the current generated by each microcell in response to the detection of a single photon is divided between the two readout pads $L$ and $R$ according to the ratio of the current divider resistors. The difference between the conductances of $R_l$ and $R_{n-l-1}$ changes proportionally to the column index $l$, and so the difference between the signals measured at the $L$ and $R$ pads is proportional to the $x$ coordinate of the microcell with respect to the LG-SiPM active area.

The horizontal buses work equivalently to the vertical ones - bringing the signal to the $T$ and $B$ readout pads, which then provide information about the $y$ coordinate of the fired microcell. For each vertical and horizontal bus, the parallel of the two current divider resistors is kept constant with respect to the column or row index, so that - to a first approximation - each column or row of microcells sees the same impedance towards the readout pads. This allows the horizontal and vertical buses to operate independently, with each one delivering only half of the charge generated by the triggered microcell. In this way, pincushion distortions, which are typical of some other types of position-sensitive devices, are avoided. Two capacitors are also added in parallel to the current divider resistors to improve the dynamic response of the detector.

\subsection{LG-SiPM Application for Optical TPC Readout}
\label{subsec:lgSiPM_application}

\begin{figure}[ht]
\centering
\vspace{5mm}
\includegraphics[width=0.95\textwidth]{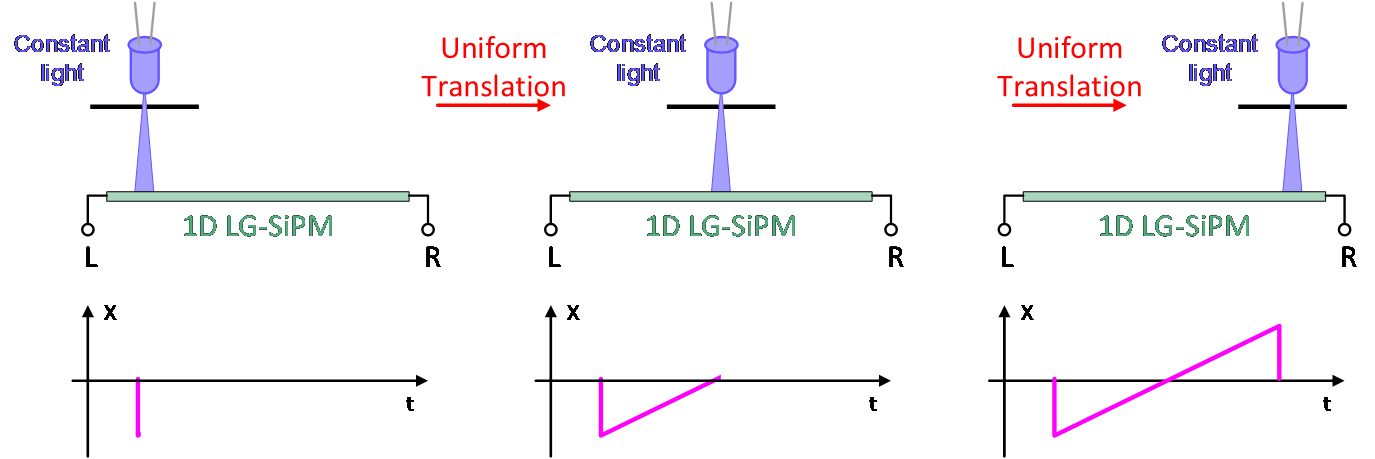}
\caption{\label{fig:SiPM-Motion}Schematic of the setup used to verify the response of the LG-SiPM to a moving light source. The progression of the light spot's $x$ coordinate, given by Equation~\ref{equ:xDistribution} at any single point in time, is shown by the waveform. The centre of the LG-SiPM is taken as the origin of the coordinate system.}
\end{figure}

Consider a simplified case of a single photon incident on a single microcell. The $x$ and $y$ coordinates and the total charge $Q$ of the microcell are related to the $T$, $B$, $L$ and $R$ readout signals by the following:

\begin{equation}
\label{equ:totalCharge}
\begin{split}
\frac{Q}{2} & = L + R \\
 & = T + B
\end{split}
\end{equation}

\begin{equation}
x = \frac{R-L}{R+L}
\label{equ:xDistribution}
\end{equation}

\begin{equation}
y = \frac{T-B}{T+B}
\label{equ:yDistribution}
\end{equation}
\\
\noindent In a more realistic scenario of a non-point-like light source incident on multiple microcells, equations~\ref{equ:xDistribution} and \ref{equ:yDistribution} now respectively describe the $x$ and $y$ coordinates of the centroid of the distribution, and $Q$ is now also proportional to the number of incident photons. As the signals generated by each microcell are combined at the four readout pads, it is no longer possible to obtain information about the individual photons that make up the light distribution.
\\
\\
For a moving light source, the equations of motion - $x(t)$ and $y(t)$ - are simply given by the variation of equations~\ref{equ:xDistribution} and \ref{equ:yDistribution} in time. This has been verified experimentally, using a simplified setup comprising a light source collimated to a 1~mm diameter circular spot, and a motorised translation stage, as shown in Figure~\ref{fig:SiPM-Motion}. By moving the light spot from the left side of the LG-SiPM active area to the right at a constant speed, the progression of the spot's calculated $x$ coordinate (shown by the waveform in Figure~\ref{fig:SiPM-Motion}) is found.

This setup involves a relatively slow rate of motion of the light source - limited by the translation stage's motor - compared to the passage of particles through an actual particle detector. However, as noted previously, the LG-SiPM response is on the order of tens of ns, and it can therefore track much faster movements than this setup allows for.

\begin{figure}[ht]
\centering
\vspace{5mm}
\includegraphics[width=0.95\textwidth]{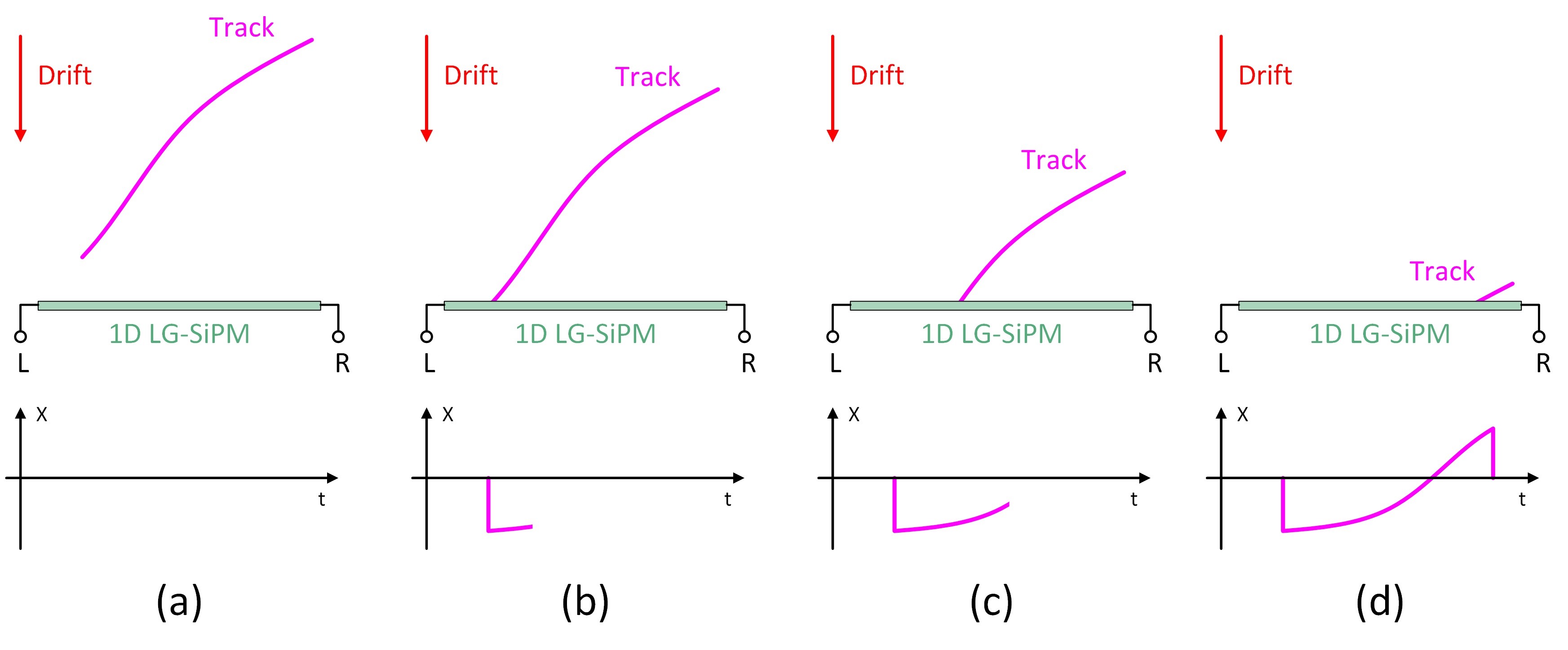}
\caption{\label{fig:SiPM-TPCConcept}The response of the LG-SiPM to an angled ionisation track in a simplified 1D TPC. The drift electrons produce secondary scintillation light in the THGEM (not shown), and this light is then incident on the LG-SiPM. The relative time and position of the incident light results in the specific duration and shape of the output waveform.}
\end{figure}

\noindent The discussion above is directly analogous to the application of the LG-SiPM to optical TPC readout: from the point of view of the LG-SiPM, the secondary scintillation light produced in the THGEM is equivalent to a time-varying light source. This is depicted in Figure~\ref{fig:SiPM-TPCConcept} for a simplified 1D TPC containing an ionisation track.

In the case of a generalised track geometry as shown, the first light is incident on the left side of the LG-SiPM (negative $x$ coordinate). The LG-SiPM then proceeds to generate a continuous analog signal as the subsequent light reaches it. The specific shape and duration of the waveform at different points in time relate to the rate and intensity of the incident light, which in turn are a reflection of the initial track.

\begin{figure}[ht]
\centering
\includegraphics[width=0.6\textwidth]{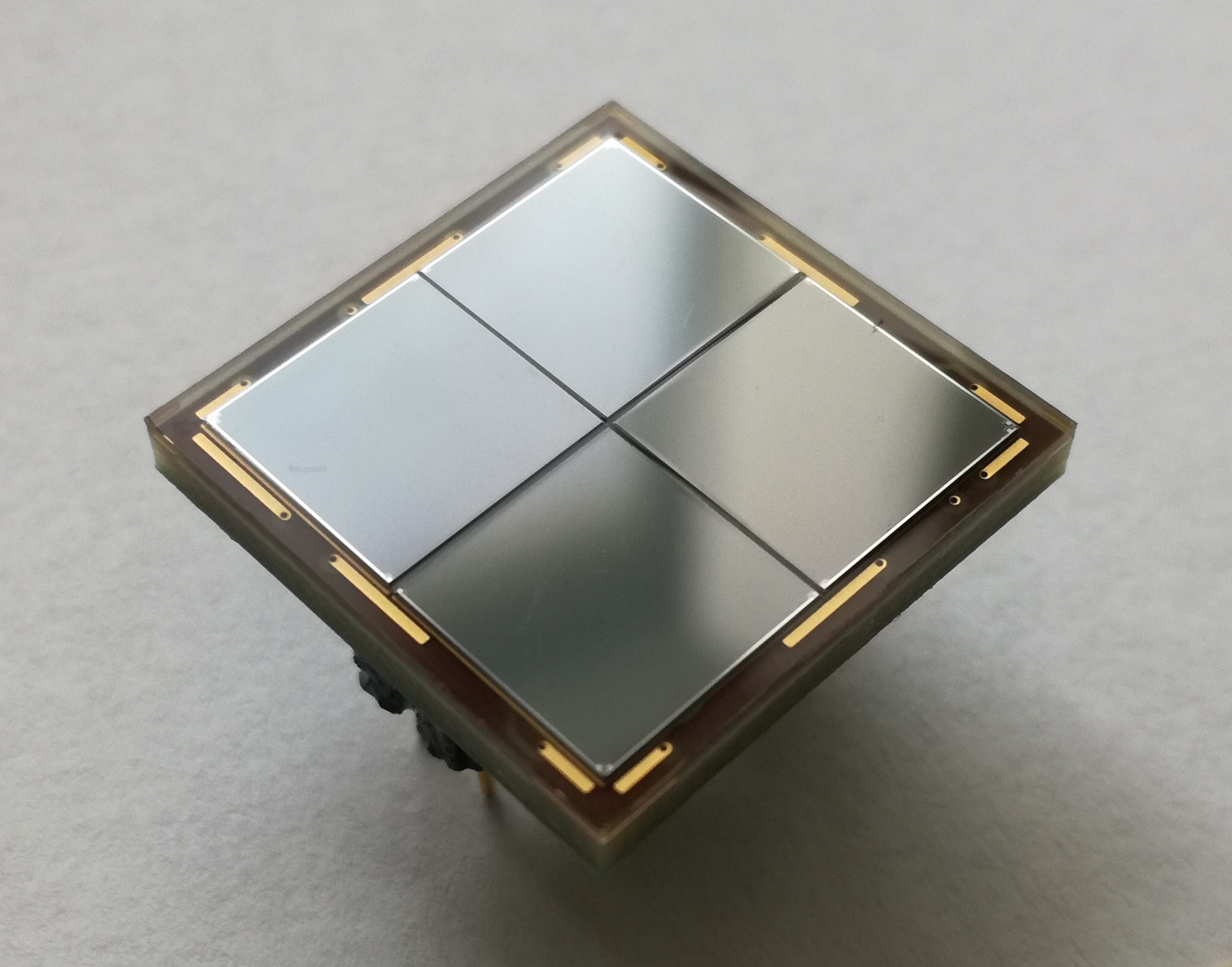}
\caption{\label{fig:LG-SiPM}The $2 \times 2$ array of LG-SiPMs used in these studies. Power and readout are provided by a separate PCB that attaches to the underside of the array via a pin-terminal.}
\end{figure}

A simple extension of this to a 2D LG-SiPM - which provides waveforms for both $x(t)$ and $y(t)$ - shows that full 3D reconstruction of an ionisation track within the TPC is possible, with the common $t$ of the waveforms being directly related to $z$ via the drift velocity of the material present in the TPC volume.
\\
\\
The results presented in this paper involved the use of a $2 \times 2$ array of LG-SiPMs, shown in Figure~\ref{fig:LG-SiPM}, which in theory would result in a total of 16 output channels (4 per device). However, a multi-die readout method has been implemented alongside the array, allowing the number of output channels to be reduced to just 6 (three per axis). This significantly decreases the complexity - both of the hardware and of the subsequent analysis - and reduces the power consumption of the readout electronics. A detailed description of this multi-die readout method will be the subject of a separate upcoming article on LG-SiPMs.

\vspace{5mm}

\noindent The spatial resolution of the track reconstruction, which is an important measure of the effectiveness of the LG-SiPM readout, is dependent on how ``clean'' the overall signal is at the point of readout. One potential factor in this is the \textit{dark count rate} (DCR) of the LG-SiPM. A dark count occurs when a microcell is fired by a carrier produced by thermal generation in silicon. It can be assumed that the DCR is a uniform distribution across the LG-SiPM's active area, and so the light signal from a track will always be superimposed on top of this flat background.

The DCR is therefore directly related to the operating temperature of the LG-SiPM, which is primarily affected by the ambient temperature. The bias voltage, $V_{bias}$, that is provided to the device also has an impact - with a higher $V_{bias}$ leading to higher temperatures in the electronics, and therefore more thermal carrier generation. However, the signal from each microcell and the photon detection efficiency also scale with $V_{bias}$, and so its value must be carefully chosen to maintain a balance between increased noise and increased signal. 

For the purposes of the work presented in this paper, the LG-SiPM was operated at room temperature with $V_{bias}$ ranging from -34 to -36~V (noted where appropriate). In this scenario, the DCR is $\approx 450$~kHz/mm$^2$, corresponding to less than 600 dark counts across the entire active area of the LG-SiPM within a 5~$\mu$s time window (this being the approximate duration of a S2 signal). When compared to the many thousands of detected S2 photons per event (as calculated in Section~\ref{subsec:detectorResponse}), it can be seen that the contribution of the DCR to the spatial resolution can be considered negligible. Even so, studies have shown that operating the LG-SiPM at cryogenic temperatures can improve the performance in term of power consumption and energy resolution by reducing the DCR to effectively zero \cite{Acerbi2017}.

\subsection{Detector DAQ and Triggering}
\label{subsec:daq}

Data acquisition and readout of the 40l TPC is performed using a bespoke software framework developed at the University of Liverpool. The PMT (1 channel) and LG-SiPM (6 channels) are all read out via a single 8-channel 12-bit 250~Ms/s CAEN V1720 digitiser, with the PMT channel additionally being branched to a threshold discriminator. A global event trigger is issued from this discriminator if the PMT (S1) signal exceeds a pre-defined threshold, and the trigger instructs the digitiser to begin reading out all 7 channels synchronously. Each event window is 16~$\mu$s long, giving 4000 samples per channel per event. Data is stored in ASCII format, with one file per channel.

\section{Results and Analysis}

\subsection{Detector Response}
\label{subsec:detectorResponse}

\begin{figure}[ht]
\centering
\includegraphics[width=0.49\textwidth]{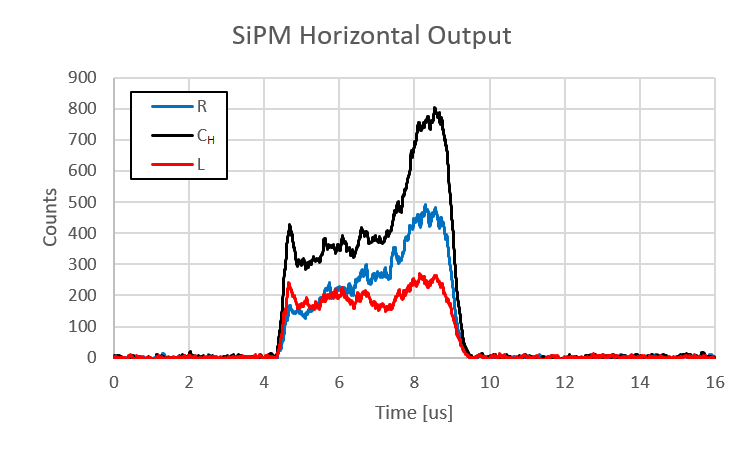}
\includegraphics[width=0.49\textwidth]{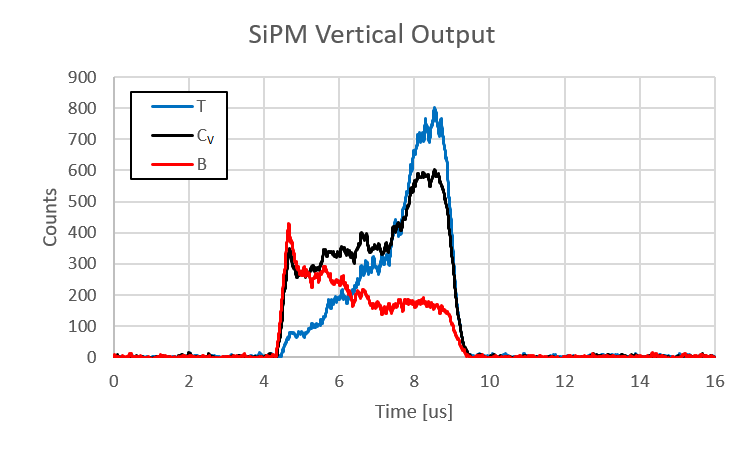}
\caption{\label{fig:sipmSignals}The $\alpha$-induced LG-SiPM signals from the $x$ axis channels $L$, $C_H$ and $R$ (left) and $y$ axis channels $T$, $C_V$ and $B$ (right). This event was recorded at a $V_{bias}$ of -34~V. The origin of the time axis corresponds to the global event trigger as discussed in Section~\ref{subsec:daq}.}
\end{figure}

As noted in Section~\ref{subsec:lgSiPM_application}, the LG-SiPM array has 6 output channels: 3 along the $x$ axis (denoted by Left [$L$], Central Horizontal [$C_H$] and Right [$R$]) and 3 along $y$ (Top [$T$], Central Vertical [$C_V$] and Bottom [$B$]). A typical set of these output channels for a single $\alpha$-induced event is shown in Figure~\ref{fig:sipmSignals}.

\begin{figure}[ht]
\centering
\includegraphics[width=0.6\textwidth]{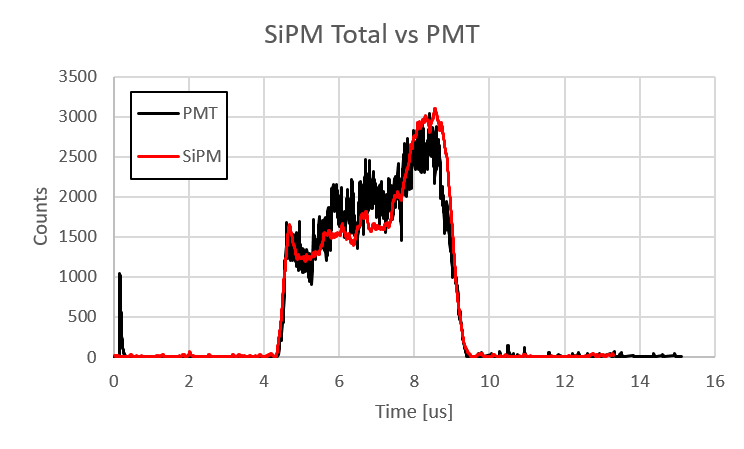}
\caption{\label{fig:pmtSignal}The summed LG-SiPM signal across all 6 channels (red) from the same $\alpha$ event as displayed in Figure \ref{fig:sipmSignals}. Also shown is the PMT signal (black), which has been scaled to the same relative peak height as the LG-SiPM signal. The signals have the same overall shape - as expected, since they both depict the same S2 scintillation light.}
\end{figure}

Figure~\ref{fig:pmtSignal} shows the PMT and total LG-SiPM signals for the same $\alpha$ event (where ``total'' refers to the summation over all 6 channels). The PMT signal has been scaled so as to have the same relative peak height as the LG-SiPM signal, allowing a clearer comparison between the signal shapes. It can be seen that the PMT detected the S1 signal at time ``zero'', issuing the global event trigger and beginning the 16~$\mu$s recording window for this event. The shapes of the PMT S2 and total LG-SiPM signals are very similar, as expected. The number of photons detected by the LG-SiPM can be determined using the total integral of the scintillation signal together with the known gain of the device (which combines the conversions from ADC counts to charge and from charge to number of photons). In this typical event, this number is calculated to be $\approx$ 5000 photons.

\subsection{Single Event 2D Track Reconstruction}
\label{subsec:track_recon_2D}

Following from the discussion in Section~\ref{subsec:lgSiPM_application}, the $x-y$ projection of a single particle track can be found by considering the relative, and time-varying, ratios of the readout channels. In the case of 3 readout channels per axis rather than 2, equations~\ref{equ:totalCharge} to \ref{equ:yDistribution} become:

\begin{equation}
\label{equ:totalCharge_3chan}
\begin{split}
\frac{Q}{2} & = L + C_H + R \\
 & = T + C_V + B
\end{split}
\end{equation}

\begin{equation}
x = \frac{R-L}{R+C_H+L}
\label{equ:xDistribution_3chan}
\end{equation}

\begin{equation}
y = \frac{T-B}{T+C_V+B}
\label{equ:yDistribution_3chan}
\end{equation}
\\
\noindent with each letter denoting the value of the signal on each corresponding channel at any given moment in time.

Due to the intrinsic variability of the digitiser as well as photon dispersion within the detector, it is impractical to calculate the $x$ and $y$ coordinates for every single datapoint covered by the S2 signal. However, using time ``slices'' representing many points at once helps to reduce the point-to-point variation considerably. The $x-y$ reconstruction is therefore determined as a series of points, with each one representing a 80~$ns$ (20 datapoint) wide time slice across all 6 channels, and its $x$ and $y$ coordinates calculated using equations~\ref{equ:xDistribution_3chan} and \ref{equ:yDistribution_3chan}. The values of $L$, $C_H$, $R$, $T$, $C_V$ and $B$ are the respective mean values within the time slice.

\begin{figure}[ht]
\centering
\vspace{3mm}
\includegraphics[width=0.49\textwidth]{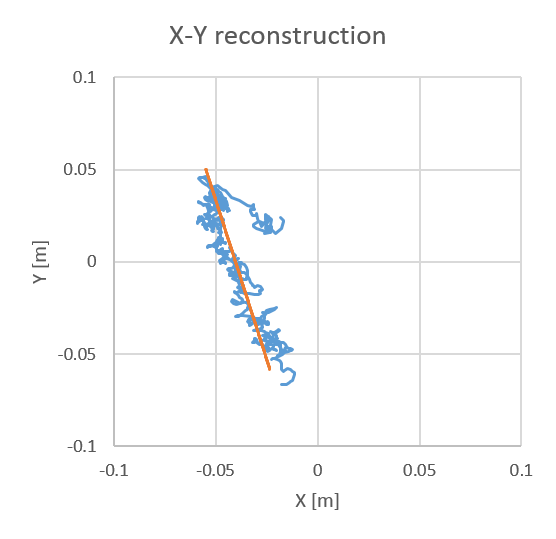}
\includegraphics[width=0.49\textwidth]{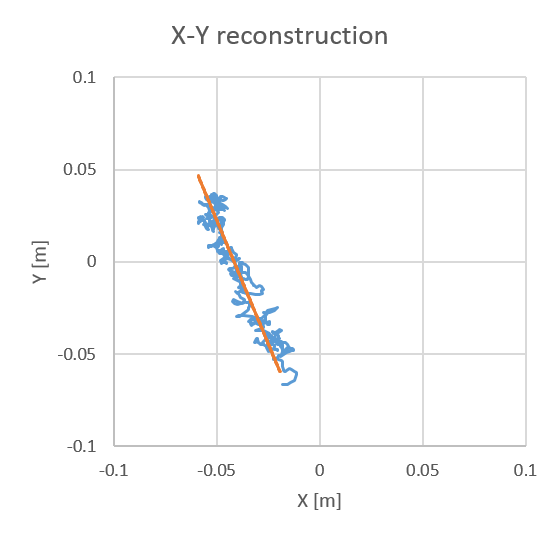}
\caption{\label{fig:xyReconstruction}(Left) The uncut $x-y$ reconstruction for the $\alpha$ event displayed in Figure~\ref{fig:sipmSignals} using the method described in the main text. (Right) The reconstruction after performing a cut on the LG-SiPM recharge component. In both cases, the line indicates a linear fit to all points using the Least Squares Method.}
\end{figure}

Figure \ref{fig:xyReconstruction} (left) shows the results of this for the $\alpha$ event previously displayed in Figure~\ref{fig:sipmSignals}.
\\
\\
Using the time ordering of the points, the general direction of the $x-y$ projection is determined to be from low to high $y$ and high to low $x$. This is consistent with the known orientation of the $\alpha$ source relative to the LG-SiPM field of view.  However, it can be seen that there are a number of points at the end of the projection that do not follow the linear trend, but instead appear to point back towards the centre of the LG-SiPM. This is a feature common to all events, and is due to the non-zero ``recharge time'' of the LG-SiPM microcells.

Each microcell require approximately 150~ns to discharge and reset after charging up due to an incident photon, and so there will be some charged microcells continuing to send signals for a period of time after the last photons from an event are incident on the LG-SiPM. This means that the last few time slices used in the $x-y$ reconstruction (the same ones that form the anomalous section in Figure~\ref{fig:xyReconstruction} (left)) are in fact non-physical and should not be considered in the reconstruction.

Figure~\ref{fig:xyReconstruction} (right) shows the $x-y$ reconstruction when the last 600~ns of the LG-SiPM signal is not used. It can be seen that the points now form a more linear distribution with a clear single direction.

\subsection{Single Event 3D Track Reconstruction}
\label{subsec:track_recon_3D}

Under the general operating principle of TPCs, the time between the S1 pulse and any point on the S2 pulse corresponds to the time taken for the free electrons at that corresponding point along the track to drift up to the THGEM. This is related to the depth ($z$ coordinate) of that point on the track through the drift velocity, $v_d$, of the scintillator medium at the particular drift field gradient and fluid pressure being used. For 100~mbar CF$_4$ at a gradient of 95~V/cm, this velocity is on the order of tens of mm/$\mu$s \cite{cf4_driftVel}.

From Figure~\ref{fig:pmtSignal} and the discussion in Section~\ref{subsec:daq}, the S1 pulse occurs at time ``zero'' in each event, and so the $z$ coordinate at time $t$ is simply given by:

\begin{equation}
z_t = v_d \times t
\label{equ:zCoordinate}
\end{equation}
\\
\noindent As with the $x-y$ reconstruction, calculating the $z$ coordinate of every datapoint in the S2 pulse will result in an impractically noisy track due to the underlying point-to-point variability of the data. Instead, the same time-slicing method is employed, using the same slice width as above and representing each slice using its central time value.

\begin{figure}[ht]
\centering
\vspace{3mm}
\includegraphics[width=0.49\textwidth]{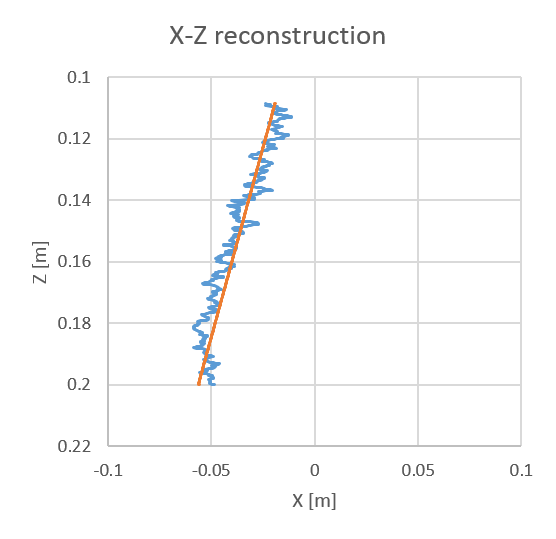}
\includegraphics[width=0.49\textwidth]{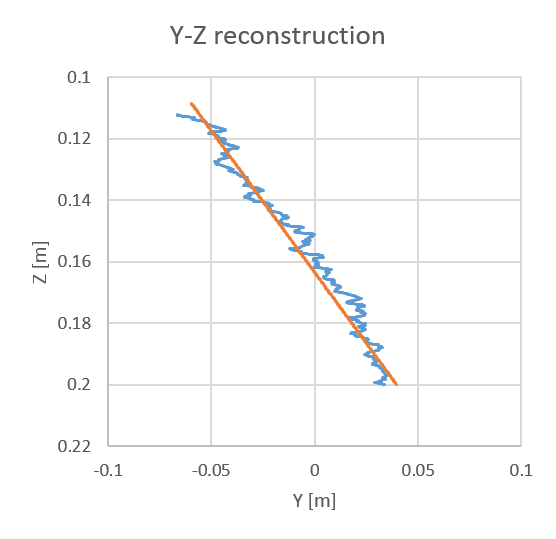}
\caption{\label{fig:xzyzReconstruction}The $x-z$ (left) and $y-z$ projections of the 3D track reconstruction, after a cut has been made based on the LG-SiPM recharge time. In both cases, the line is a linear fit to all points using the Least Squares Method.}
\end{figure}

The $x-z$ and $y-z$ projections of the resulting 3D track reconstruction are shown in Figure~\ref{fig:xzyzReconstruction}. As with the $x-y$ reconstruction, a cut on the last 600~ns of the S2 pulse (discussed previously in Section~\ref{subsec:track_recon_2D}) has been applied. Although there is still some noticeable point-to-point variation, the overall downward angle of the track is consistent with the known orientation of the $\alpha$ source. The absolute $z$ values are also correct for the known position of the source relative to the top of the TPC (which is taken as $z = 0$~mm).

\subsection{Multi-Event Analysis}
\label{subsec:multiEvent_analysis}

\begin{figure}[ht]
\centering
\vspace{5mm}
\includegraphics[width=0.49\textwidth]{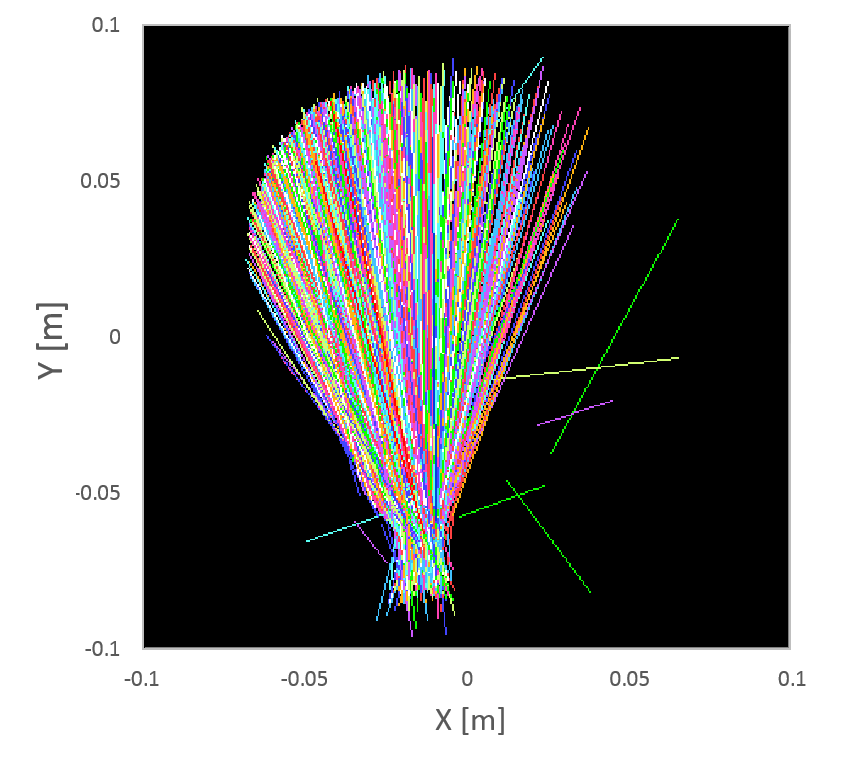}
\includegraphics[width=0.49\textwidth]{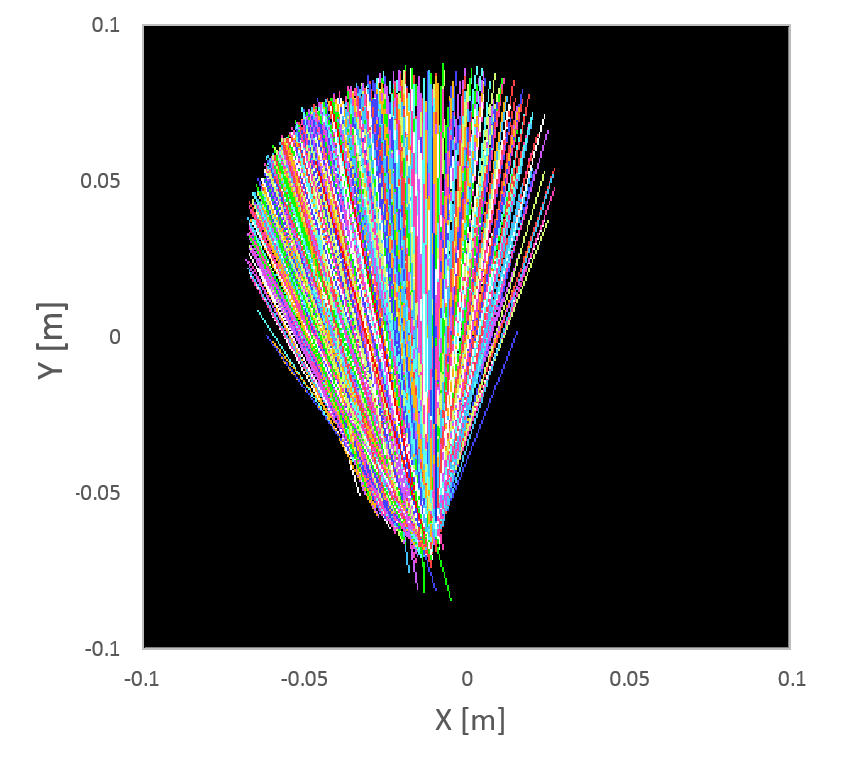}
\caption{\label{fig:xy_allTracks}(Left) The overlaid $x-y$ projections of all reconstructed tracks in a single run, taken at a $V_{bias}$ of -36~V and a 60{\textdegree} $\alpha$ source angle. (Right) The same overlay for only the filtered events - i.e. after applying cuts on the PMT S2 signal area and reconstructed 3D track length.}
\end{figure}

\begin{figure}[ht]
\centering
\vspace{3mm}
\includegraphics[width=0.8\textwidth]{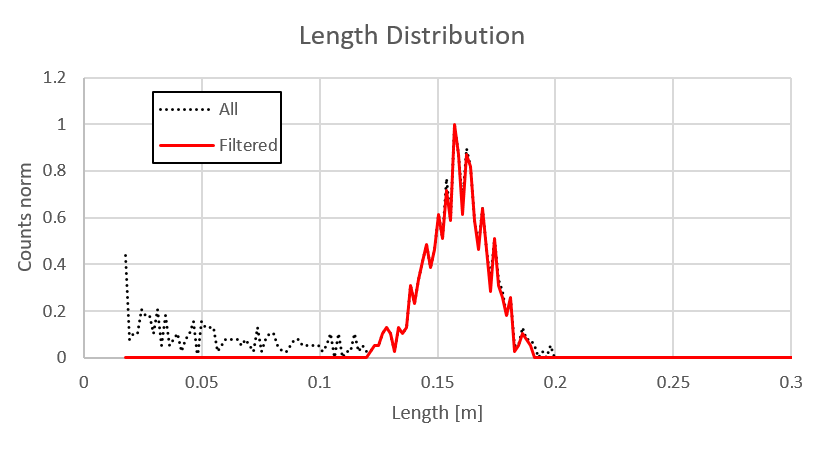}
\caption{\label{fig:trackLengths}(Black) The (normalised) distribution of reconstructed 3D track lengths across all events in the same run depicted in Figure~\ref{fig:xy_allTracks}. (Red) The same distribution for only the filtered events - i.e. after applying cuts on the total PMT S2 signal area and reconstructed 3D track length.}
\end{figure}

The overlaid $x-y$ projections of all reconstructed tracks in a single run is shown in Figure~\ref{fig:xy_allTracks} (left). A cone of tracks can clearly be seen, with an opening angle and forward direction consistent with the shape and orientation of the $\alpha$ source and its collimator. Figure~\ref{fig:trackLengths} (black) shows the distribution of 3D track lengths across all reconstructed tracks. The vast majority of tracks have lengths between 150 and 200~mm - as expected from a mono-energetic, single-species particle source embedded in a homogeneous scintillator medium.

It can be seen that there are a small number of tracks with much shorter lengths than the majority. It was found that these events also have significantly smaller S2 pulses than expected (in terms of both duration and total area), suggesting that they are of much lower energy than the rest of the events. The small S2 pulses make it difficult to accurately reconstruct the tracks - as can be seen on Figure~\ref{fig:xy_allTracks} (left), where many of these short-length tracks appear to go backwards into the collimator.

\begin{figure}[ht]
\centering
\vspace{3mm}
\includegraphics[width=0.8\textwidth]{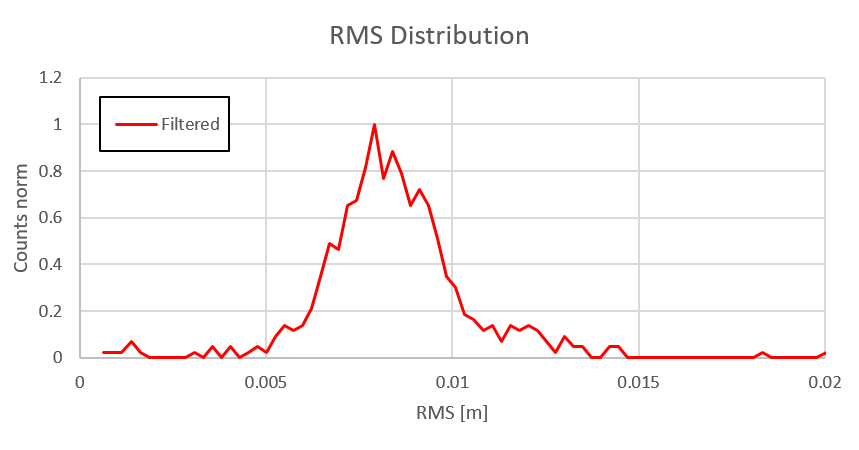}
\caption{\label{rmsDistribution}The (normalised) distribution of the $x-y$ reconstruction error across the filtered events. This error is calculated (per event) as the RMS of the distance between each calculated $(x, y)$ point and the linear fit to all points.}
\end{figure}

It is presumed that such events are $\alpha$ particles that have become ``caught'' in the collimator after being emitted, subsequently losing a significant fraction of their energy due to scattering and/or other interactions. Such events are not representative of the LG-SiPM characterisation being studied here, and can be removed by applying cuts on the PMT S2 signal area and reconstructed 3D track length. The effect of these cuts is shown in Figure~\ref{fig:xy_allTracks} (right) and Figure~\ref{fig:trackLengths} (red). For convenience, the events remaining after these cuts are referred to hereafter as the \textit{filtered} events.
\\
\\
The uncertainty in the $x-y$ track reconstruction of a single event has been quantified as the RMS of the distance between each calculated point in the $x-y$ plane and the linear fit to all points. Figure~\ref{rmsDistribution} shows the distribution of the RMS across the filtered events. For this particular run, the mean error on the $x-y$ track reconstruction is $\approx$ 8~mm.

\begin{figure}[ht]
\centering
\vspace{5mm}
\includegraphics[width=0.49\textwidth]{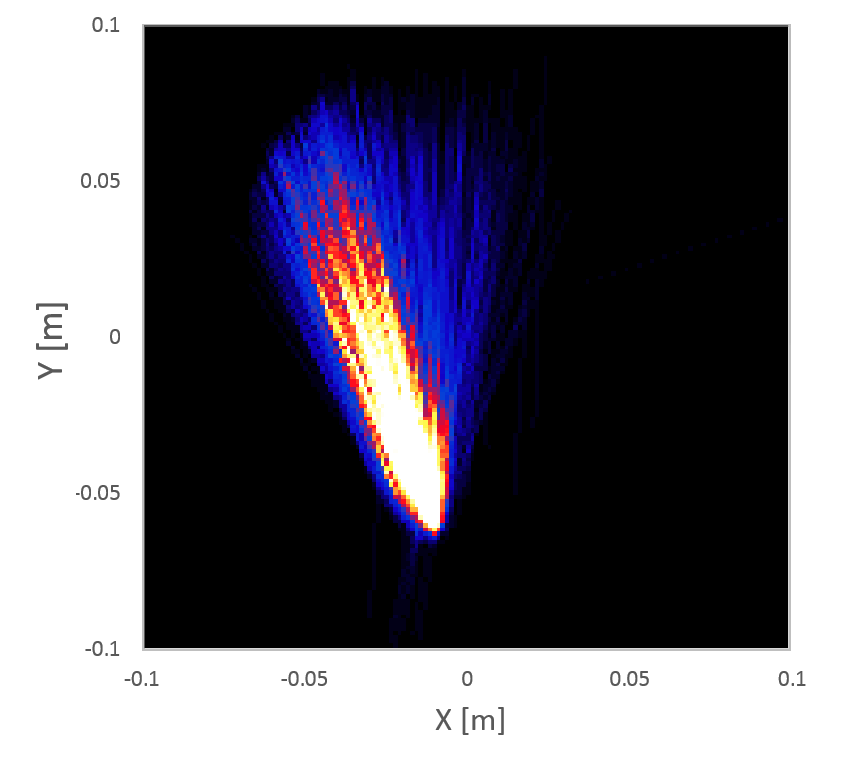}
\includegraphics[width=0.49\textwidth]{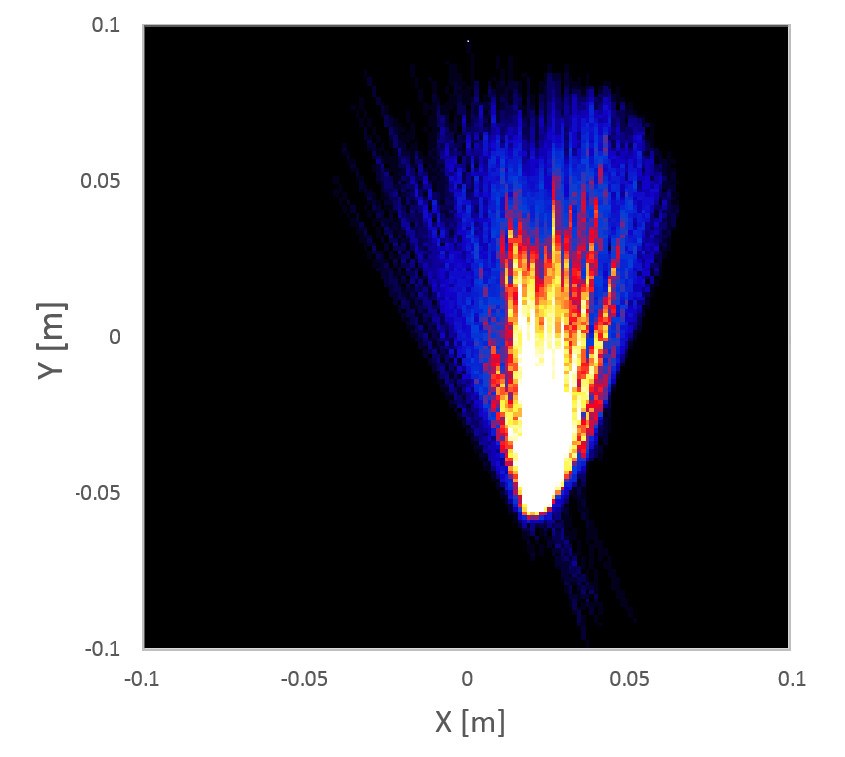}
\caption{\label{fig:density_allTracks}The track density distribution for an $\alpha$ source angle of 60{\textdegree} (left) and 45{\textdegree} (right) across datasets of filtered events. White colour indicates the highest density, and blue indicates the lowest. These data were collected at a $V_{bias}$ of -35~V.}
\end{figure}

Extension of this method to 3D - using the RMS of the distance between each calculated point in $(x, y, z)$ and a 3D linear fit - is possible, in order to quantify the uncertainty in the overall 3D track reconstruction. Work on this is still ongoing.
\\
\\
The directionality of the reconstructed tracks can be emphasised by considering the track density, shown in Figure~\ref{fig:density_allTracks}. The data displayed were taken at two different orientations of the $\alpha$ source: 60{\textdegree} (left) and 45{\textdegree} (right). The difference between the two distributions is evidently clear, demonstrating the good position sensitivity and resolution of the LG-SiPM readout.

\subsection{Energy Reconstruction and Deposition Rate}
\label{subsec:energy_recon}

As noted in Section~\ref{subsec:lgSiPM_application}, at a given $V_{bias}$ the total charge $Q$ produced by the LG-SiPM (given in ADU by equation~\ref{equ:totalCharge_3chan}) is directly proportional to the number of incident photons. This in turn is related to the number of electrons produced through the incident particle's ionisation of the scintillator. For the single-species source used in these studies, the amount of ionisation - and therefore $Q$ - is only dependent on the particle energy.

\begin{figure}[ht]
\centering
\vspace{3mm}
\includegraphics[width=0.8\textwidth]{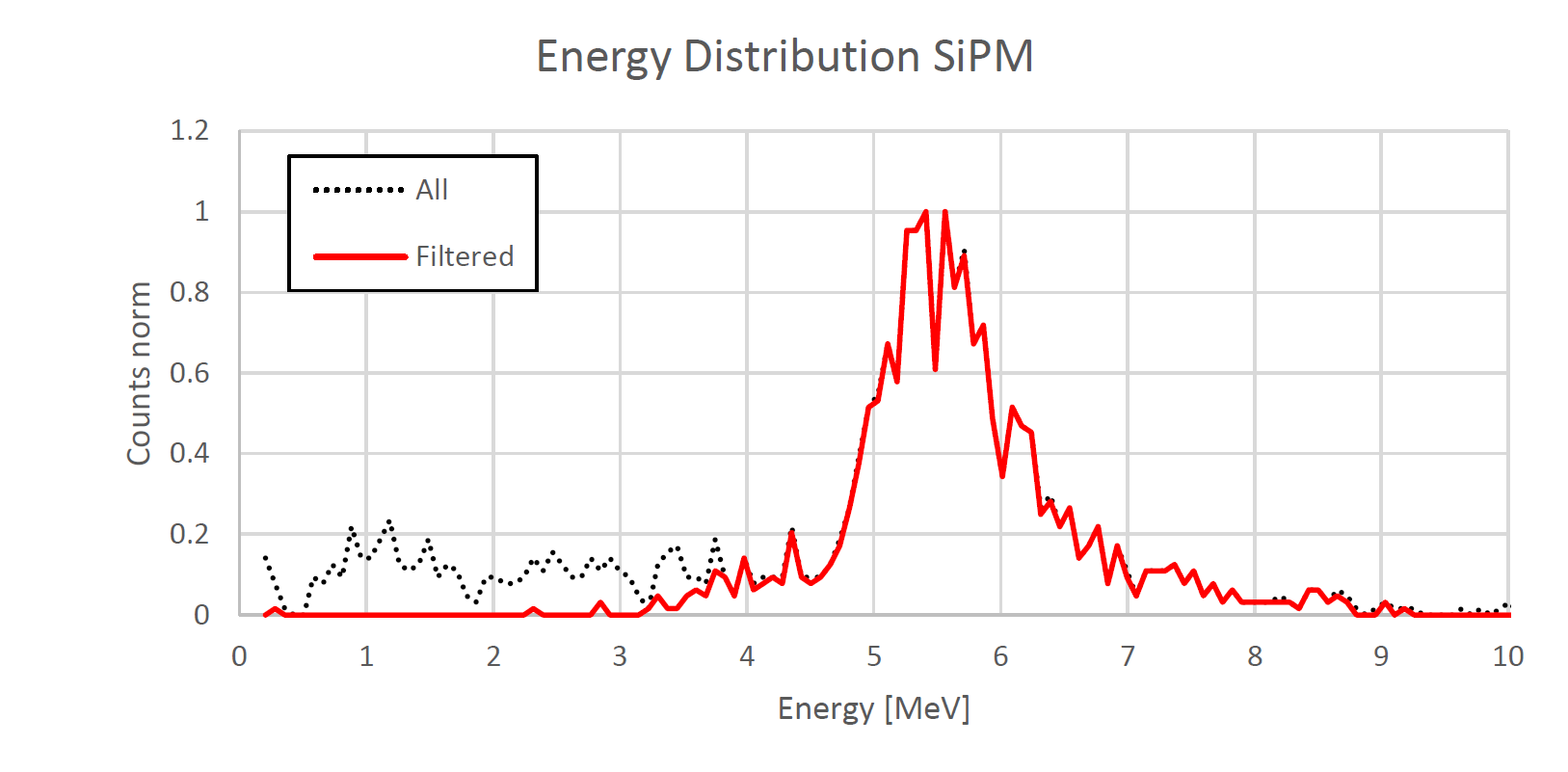}
\caption{\label{fig:energyReconstruction}The distribution of reconstructed energy (red) across filtered events in a run taken at $V_{bias} = -34~$V. This distribution is that of the total LG-SiPM charge $Q$, given by equation~\ref{equ:totalCharge_3chan}, with the peak scaled to the energy of $^{241}Am$ $\alpha$ particles. For comparison, the black line indicates the distribution using all events in the run, showing the low energy events that are presumed to be $\alpha$ particles that were ``caught'' in the collimator.}
\end{figure}

In principle, $Q$ should take a single value, since the $^{241}Am$ $\alpha$ particles are mono-energetic at 5.486~MeV. However, in practice, detector effects and non-zero calorimetry resolution will broaden the $Q$ values into a distribution. Nonetheless, once this distribution of $Q$ is found, a simple ADU-to-energy calibration can be used to scale the most probable $Q$ value to the expected single $\alpha$ energy. The distribution of this reconstructed energy across many events is shown in Figure~\ref{fig:energyReconstruction}.

Based on the distribution's Gaussian shape with a FWHM of $\approx$ 1~MeV, the energy resolution can be calculated as $\approx$ 42\%. While this is quite large, a number of independent sources of uncertainty are combined into this value - such as potential effects stemming from outgassing within the detector (which will affect the CF$_4$ purity over time and increase the event-to-event variability of the number of electrons produced during ionisation), THGEM instabilities (which can affect the number of S2 photons produced), and the previously noted effect of the LG-SiPM's DCR (which could increase the $Q$ and contribute to the tail at higher energies). Further analysis to characterise and separate these effects is ongoing.
\\
\\
The ratio of an event's reconstructed energy to its previously reconstructed 3D track length gives the energy deposition rate: $\frac{dE}{dX}$. This quantity, which is also known as the ``Linear Energy Transfer'' (LET), is important for particle identification, as it is primarily a function of only the particle mass (and therefore, species) and momentum (which, in many particle detectors, is known or can be relatively easily determined).

\begin{figure}[ht]
\centering
\includegraphics[width=0.8\textwidth]{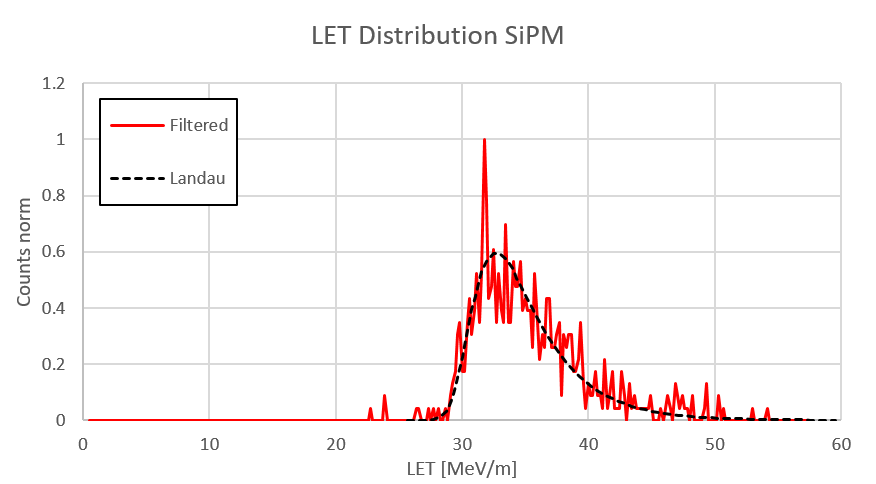}
\caption{\label{fig:letDistribution}The distribution of the Linear Energy Transfer (LET) across filtered events, with a Landau function fitted to the distribution (black). The events in this data are the same as those presented in Figure~\ref{fig:energyReconstruction}.}
\end{figure}

The distribution of the calculated LET is shown in Figure~\ref{fig:letDistribution}. It is seen to follow the expected Landau-shaped behaviour, with a most probable value at $\approx$ 32.5~MeV/m.

\section{Conclusions and Future Development}

The current LG-SiPM has shown great potential for optical readout of TPCs - having been able to successfully reconstruct the 3D tracks of individual particles to a reasonable accuracy, and demonstrating good position resolution and a signal-to-noise ratio even when operated in room temperature conditions and at a minimal $V_{bias}$. However, these studies have brought to light certain limitations with the current designs, which provide possible routes for future developments.
\\
\\
Due to the way that the information from individual microcells is combined (and therefore lost) at the readout pads, the LG-SiPM is unable to correctly reconstruct events where a large number of microcells are fired simultaneously. The two scenarios in which this was found to be particularly problematic are horizontal tracks and pileup events (where two or more tracks are present in the same event window). In the case of horizontal (and near-horizontal) tracks, all of the S2 photons arrive at the LG-SiPM at the same time, and so this is seen as a short, sharp signal in each readout axis instead of one with a flat shape and long duration (as would be seen by the PMT). The related situation of pileup and near-simultaneous events sees the individual tracks merged into a single light distribution, with no clear separation between them that would allow their individual reconstruction. This imposes an upper limit on the event rate that the LG-SiPM is able to cope with. This maximum rate is naturally also dependent on the duration of each event, and in these studies with S2 signals lasting a few $\mu$s, it is found to be approximately 100~kHz. While this is fast enough for reliable operation with low activity particle sources and cosmic rays, the current LG-SiPM design would not be suitable for use in a beamline or other high particle rate operation.

As noted, both of these limitations arise from the design of the LG-SiPM itself - simply because it is not a pixel detector, but instead combines the information across many pixels (microcells). One possible avenue for improvement to address this could be to add independent readout channels for specific regions of the active area. For example, if each quadrant of the 2 $\times$ 2 array used in these studies had its own individual $x$ and $y$ readouts alongside the 6 combined channels, signals could be spatially localised in order to provide a more detailed picture of the track geometry. (It is understood, of course, that the ultimate limit of this idea is when each microcell has its own individual readout channel, at which point the LG-SiPM effectively becomes a pixel detector.)
\\
\\
As mentioned in Section~\ref{subsec:lgSiPM_application}, the LG-SiPM's DCR would be greatly reduced by operation at cryogenic temperatures. However, the current design's electronics are not cryogenic-compatible.

Partially in response to this, a new production cycle of LG-SiPMs using FBK's NUV-HD cryogenic technology is already underway. These detectors - which are expected to have a DCR of almost 0~Hz based on initial internal testing - will be able to stably operate at cryogenic temperatures, thus benefiting from reductions in the background power consumption and thermal variances in the measurements. For the purposes of optical TPC readout, this will be reflected in less noisy output signals and smaller uncertainties in position and energy reconstruction.

An additional benefit of the NUV-HD technology is a shift in the wavelength response of the LG-SiPM. As previously noted in Figure~\ref{fig:LG-PDE}, the current device has a peak detection efficiency at 550~nm, which lies between the visible and UV secondary scintillation emissions of CF$_4$, which peak at 625 and 300~nm respectively. However, initial studies of the NUV-HD technology being used in the newest iteration of the LG-SiPM show a peak detection efficiency at 420~nm, with relatively high response at even lower (VUV) wavelengths, therefore making it much more sensitive to UV scintillation emissions - not just of CF$_4$, but also other commonly used scintillators such as liquid argon. 
\\
\\
Alongside these large-scale changes to the underlying LG-SiPM design and manufacturing, small-scale fine-tuning of the components - such as capacitor and resistor values - can be of some additional benefit. With these changes, it may be possible to reduce the impact of the recharge time and other artifacts, thereby reducing uncertainties on the position reconstruction. Digital post-processing of the readout signals can also potentially be implemented, either inline with the readout itself or as a pre-analysis software step, and this would benefit the signal-to-noise ratio and point-to-point variability. Post-processing could also improve the LG-SiPM's response to pileup, allowing individual tracks to be distinguished and increasing the maximum event rate that the device is capable of operating at.
\\
\\
It is anticipated that further testing and characterisation of both the current and future LG-SiPM designs in the context of optical TPC readout will take place on the ARIADNE TPC. This detector is a 1-ton dual-phase LArTPC with a 53 $\times$ 53~cm active THGEM area, and will therefore require a correspondingly larger scale SiPM device or devices. This is expected to be achieved through a combination of the array structure and multi-die readout previously discussed in Section~\ref{subsec:lgSiPM_application}, together with optimisations to the detector structure and optics. This will provide a larger and more representative test-bed for the ongoing development of LG-SiPM-based cameras for optical TPC readout.

\FloatBarrier
\acknowledgments

The ARIADNE research program is proudly supported by the European Research Council Grant No. 677927 and the University of Liverpool.
\\
\\
The authors would like to thank the members of the Mechanical Workshop at the University of Liverpool's Physics Department and the researchers and technicians at Fondazione Bruno Kessler for their technical expertise and contributions.

\end{document}